# Growth of VO$_2$-ZnS Thin Film Cavity for Adaptive Thermal Emission

Raymond Yu[1], Bo K. Shrewsbury[2], Claire Wu[3], Harish Kumarasubramanian[3], Mythili Surendran[3,4], Jayakanth Ravichandran[1,3,4], Michelle L. Povinelli[1,2,a]

[1]*Ming Hsieh Department of Electrical and Computer Engineering, University of Southern California, Los Angeles, CA 90089, USA*

[2]*Department of Physics & Astronomy, University of Southern California, Los Angeles, CA 90089, USA*

[3]*Mork Family Department of Chemical Engineering & Materials Science, University of Southern California, Los Angeles, CA 90089, USA*

[4]*Core Center of Excellence in NanoImaging, University of Southern California, Los Angeles, CA 90089, USA*

Low-weight, passive, thermal-adaptive radiation technologies are needed to maintain an operable temperature for spacecraft while they experience various energy fluxes. In this study, we used a thin-film coating with the Fabry-Perot (FP) effect to enhance emissivity contrast (Δε) between VO$_2$ phase-change states. This coating utilizes a novel hybrid material architecture that combines VO$_2$ with a mid- and long-wave infrared transparent chalcogenide, zinc sulfide (ZnS), as a cavity spacer layer. We simulated the design parameter space to obtain a theoretical maximum Δε of 0.63 and grew prototype devices. Using X-ray diffraction, Raman spectroscopy, and Fourier Transform Infrared (FTIR) Spectroscopy, we determined that an intermediate buffer layer of TiO$_2$ is necessary to execute the crystalline growth of monoclinic VO$_2$ on ZnS. Through temperature-dependent FTIR spectroscopy measurements, our fabricated devices demonstrated FP-cavity enhanced adaptive thermal emittance.

______________________________

[a] Corresponding Author.  Electronic mail:  povinell@usc.edu

**Main Text:**

Thermal regulation is a critical process in preserving the integrity of a system and its operation. One method of regulation is through passive adaptive radiation, in which an object emits more thermal radiation when heated and less when cooled. Adaptive thermal radiation is a pivotal tool for wearable technology[1], the mitigation of building energy consumption,[2,3] and the survival of spacecraft.[4] A common method of enabling this mechanism is through volatile phase change materials (PCMs),[5] which are materials that respond to environmental stimuli with changes in lattice structure. In particular, solid-state PCMs are much smaller and lighter in weight than solid-liquid PCMs,[6] a key feature for space applications.

$VO_2$ is a well-studied, solid-state PCM for adaptive radiation. It exhibits an electronic transition from an insulator to a metal with a critical temperature, $T_c$ ~340K,[7] near room temperature. This electronic transition is accompanied by significant changes in optical properties, where its refractive index doubles, and the extinction coefficient around a wavelength of 8 μm increases by an order of magnitude.[8,9] Moreover, the reversable change from the cold-insulating state (Monoclinic) to hot-metallic state (Tetragonal - rutile) occurs over a relatively narrow temperature range, a key feature for use as an adaptive radiator.[10]

The effectiveness of an adaptive thermal emitter is characterized by the difference in total emissivity between high- and low-temperature states, $\Delta\varepsilon$.[11] $VO_2$ on a back reflector, alone, has a theoretical limitation on $\Delta\varepsilon$ because its absorption in the cold state will dominate the hot state with increasing thickness.[12] To overcome this limitation, several works have displayed devices with enhanced $\Delta\varepsilon$ based on planar Fabry Perot- (FP) cavities,[9,13,14,15,16,17] metasurfaces,[18,19,20,21] and microstructures.[10,12] Of these geometries, the planar FP-cavity structure is the most scalable for manufacturing, as it avoids the need for any lithographic patterning.

A planar FP-cavity scheme for adaptive thermal radiation contains the following elements:[22] (1) a PCM layer ($VO_2$) that provides the emission switching effect, (2) a lossless spacer layer that enhances optical interaction with the PCM layer through interference, and (3) a back reflector that redirects the emission outwards. Previous experimental works have used oxides such as $HfO_2$,[13,21] $SiO_2$,[16,20] and $Al_2O_3$[17] as the spacer layer, but they are optically limited by the strong phonon resonance in the Mid-IR wavelength and longer.[23] This presents a particular challenge, as the majority of the blackbody spectrum at 30 °C (~303 K) spans from 5 µm to 25 µm, in which oxide-based spacers result in a higher cold-state emissivity than desired. Alternatively, halides are the most transparent material from UV to LWIR but are hygroscopic and have high water solubility.[24] Additionally, their low refractive index requires an optimal spacer thickness of ~1.2-1.8µm, creating stringent mechanical constraints.[14,15] Chalcogenides such as ZnS and ZnSe, materials commonly used as thermal imaging grade optics, are highly transmissive from the MIR to LWIR with high refractive index, but $VO_2$'s thin film growth compatibility with ZnS has not been studied.

In this work, we investigate a novel growth method of $VO_2$ with a ZnS cavity spacer layer, in a FP-cavity system for adaptive thermal radiation. ZnS is a better choice for high-speed space/aircrafts over ZnSe due to lower cost, and greater mechanical strength (hardness and thermal shock).[25] Although ZnS has a shorter transparency window than ZnSe, it still has a >90% transmission at a wavelength of 25 µm and film thickness of 1 µm.[26] Further, a subset of authors have already demonstrated high quality thin films of ZnS using pulsed laser deposition (PLD).[27] To date, there have been several design works on incorporating ZnS with $VO_2$,[28,29,30] but no experimental work on direct thin film deposition. The only existing experimental works with the two materials are solution-based $VO_2$ nanoparticles,[30,31,32] not all of which demonstrated phase-change properties. In this study, we present a novel method to crystallize monoclinic PLD $VO_2$ on ZnS, at 400 °C, through an intermediate oxide buffer layer. We characterize the materials and verify the emissivity-switching behavior of several prototype devices across a wide spectral range of 2-18 µm. Promising results for $VO_2$-ZnS thin film growth demonstrated here will provide new material platforms and approaches for satisfying passive thermal control demands on Earth and in Space.

We used PLD as the growth method for the oxides and ZnS. Following existing work on epitaxial $VO_2$ growth,[33] all oxide PLD growth in this study was done at 400 °C, oxygen partial pressure of 10 mTorr, and 248 nm KrF excimer laser fluence of 1.5 J/cm$^2$. Sulfide growth was performed in a separate chamber at 400 °C, with a tert-butyl disulfide precursor gas at 1mTorr and a laser fluence of 1.0 J/cm$^2$.[27] First, we verified the growth of ~200 nm ZnS on a platinum-based back reflector, denoted as sample (s1) in Figure 1a. The powder X-ray diffraction (XRD) pattern is plotted as the blue data set in Figure 1b. The Pt film had a strong preferred 111 orientation with weak texture along 00$l$ and subsequently, the ZnS layer only shows 002 and 004 reflections suggesting 00$l$ type out-of-plane texture. Here, we presume that the ZnS is in the wurtzite structure as reported,[27] and 111 oriented Pt could aid the crystallization of wurtzite over sphalerite phase. In Figure 1c, the sample did not exhibit any Raman peaks, as expected from exciting below the bandgap (3.6 eV).[34]

We next verified the growth of crystalline $VO_2$ by direct deposition on to the back reflector, which yields sample (s2) with a corresponding data set color of orange. We also incorporated a thin layer of $HfO_2$ to passivate the surface of the $VO_2$,[35] while avoiding significant alteration in the optical property of our prototype adaptive emitter. The measured thicknesses of the $HfO_2$ and $VO_2$ layers using X-ray reflectivity (XRR) were ~20 nm and ~60 nm, respectively. All our PLD grown films exhibit polycrystalline structure but with a preferred texture, given the textured nature of the Pt metal layer. The X-ray diffraction (XRD) patterns of Sample (s2) are depicted in Figure 1b, revealing broad diffraction peaks at approximately ~28.1° and ~46.8° corresponding to $VO_2$ 111 and $HfO_2$ 111 reflections. The predominant XRD signal arises from $VO_2$, which crystallizes first and grows thicker. Additionally, we obtained Raman shift spectra, as illustrated in Figure 1c, and observed spectra are consistent with literature values.[33]

Our proposed adaptive thermal emitter, sample (s3) was grown simultaneously as (s2), using (s1) in place of a plain back reflector. From the XRD pattern, shown by the red data set, it is difficult to distinguish the contribution of VO$_2$ from ZnS patterns. However, the observation of only one matched Raman peak in Figure 1c clearly indicates the lack of crystallization of monoclinic VO$_2$, when grown directly on ZnS at 400 °C. The material is thus labeled as x-VO$_2$ in Figure 1a.

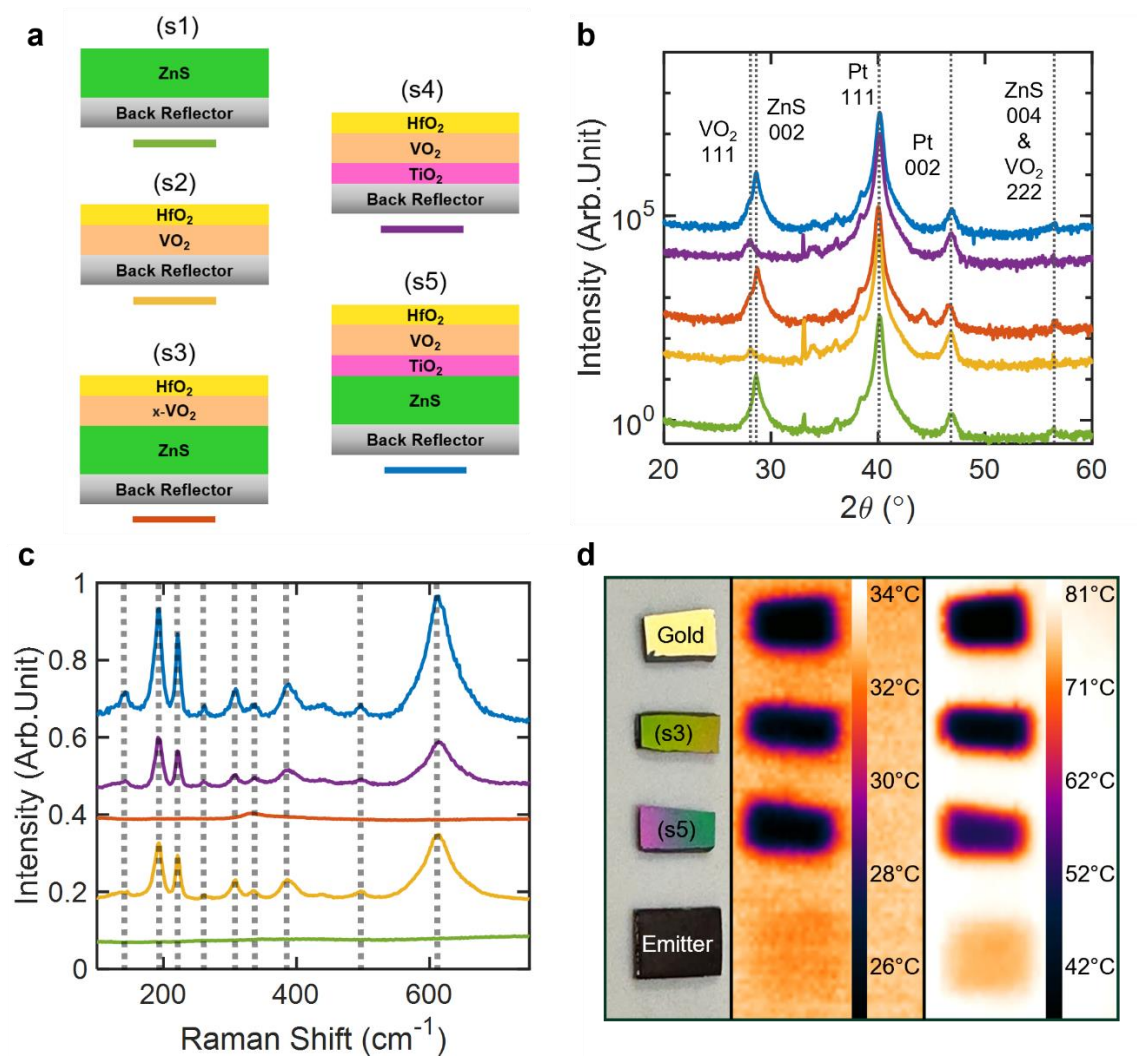

**FIG. 1.** (a) Schematic depicting several stack combinations of ZnS, VO$_2$, HfO$_2$, and TiO$_2$ on a metallic back reflector, (b) XRD patterns for each stack combination, (c) Raman spectra (excited at 532 nm) with dashed lines representing literature VO$_2$ peaks,[33] (d) The left column shows optical images of a gold reference sample, (s3), (s5), and a reference thermal emitter. The middle and right columns show thermal camera images taken in cold (~32°C) and (~80°C) hot- states, respectively, where the color bar indicates apparent temperature measured by the camera.

To facilitate the crystallization of $VO_2$ on ZnS, we used $TiO_2$ as a buffer layer to reduce interfacial energy. Rutile is the most stable polymorph of $TiO_2$.[36] As the high temperature phase of $VO_2$ shares the same rutile structure with similar lattice parameters,[37] it is widely used as an epitaxial substrate for growing $VO_2$. Hence, we anticipate that a $TiO_2$ buffer layer will promote nucleation of $VO_2$ over $HfO_2$. To evaluate our hypothesis, we grew sample (s4) and (s5), simultaneously, under the same laser fluence, oxygen partial pressure, and temperature conditions as (s2) and (s3). The Raman spectra in Figure 1c clearly show that both (s4) and (s5) have reflections matching those of monoclinic $VO_2$ (dashed lines).

Furthermore, we observed direct evidence of a thermal emissivity change in (s5). We compared a low-emissivity, gold reference sample, (s3), (s5), and a paper-based, high-emissivity reference sample. We recorded images of the samples on a hot plate, using a Seek Thermal camera. The results are shown in Figure 1d for both a cold state (30 °C) and a hot state (80 °C). The color bars indicate the apparent temperature as measured by the camera. Initially, at low temperature, sample (s5) presents the same color as the gold reference sample and (s3). As the samples were heated, (s5) became brighter in color than gold and (s3). This indicates a larger relative change in thermal emission with respect to the cold state.

Since our primary method of material growth is PLD, the refractive index and extinction coefficient can differ from existing literature values of other growth methods[8,38] and affect our emissivity calculations. In Figure 2a & 2b, we present the measured $n$ and $k$ of $VO_2$ insulating state (I), $VO_2$ metallic state (M), and ZnS measured using Angstrom-Sun Technologies Inc. Infrared Spectroscopic Ellipsometer, in the 2-15 μm range. The optical constants were fitted through a commercial software, TFProbe and we extrapolated the $n$ and $k$ values up to 18 μm to match the FTIR spectral range. The optical constants of $HfO_2$ and $TiO_2$ in Figure 2a and 2b were provided by the ellipsometer database,[39] as small variations in the optical constants of ~20 nm dielectric films will minimally impact device performance. From Figure 2a and 2b, we observe that ZnS has a low dispersion and is also incredibly transparent, with $k$ values in the ~$10^{-3}$ range. Next, we observe that $VO_2$(I)'s refractive index and absorption coefficient drastically increased as it phases change into the metallic state (VO2(M)). We also observe a high $k$ in the LWIR for the oxide materials (HfO$_2$, VO$_2$(I), TiO$_2$), due to the presence of phonon resonances.

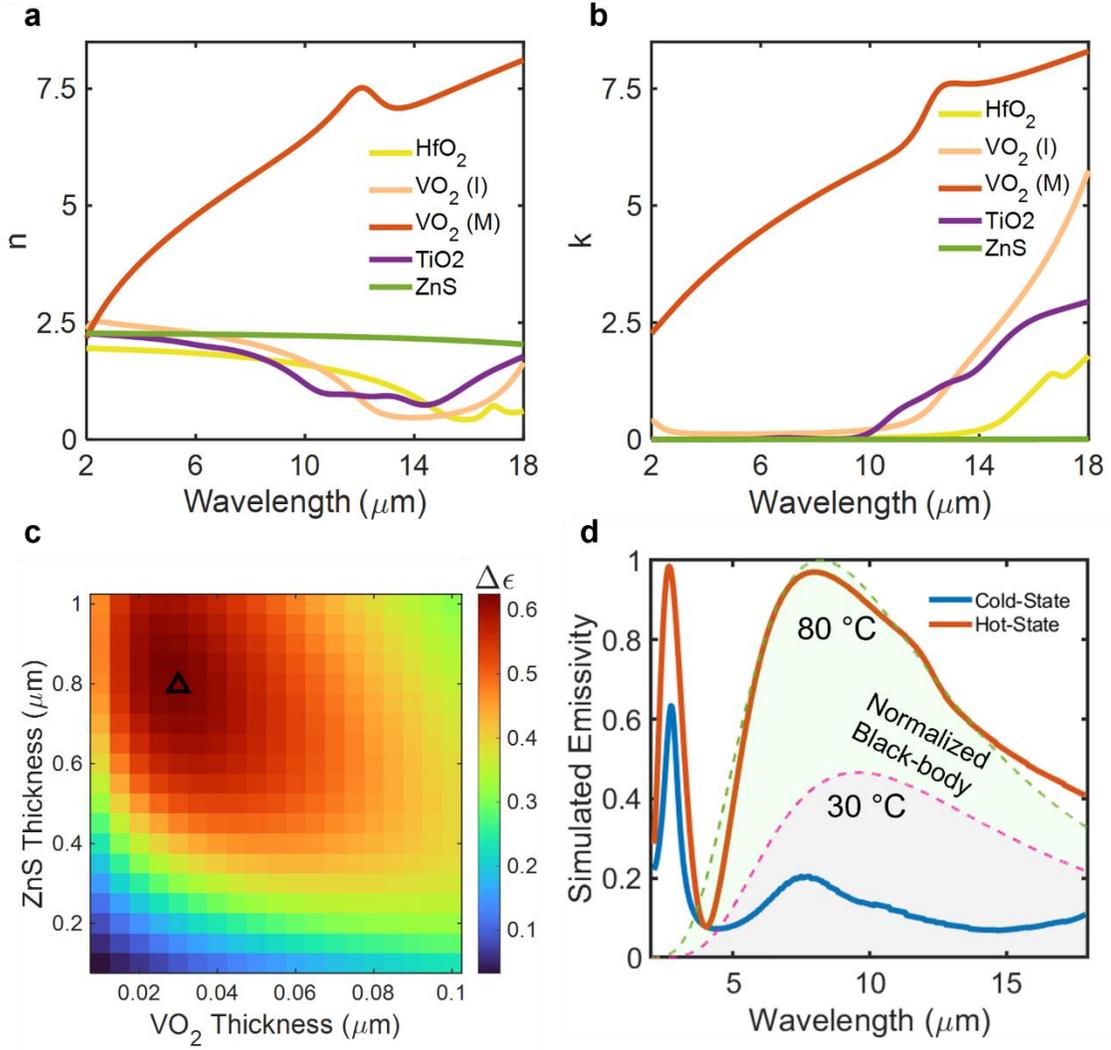

**FIG. 2.** The (a) refractive index and (b) extinction coefficient of the materials in this study; VO2(I), VO2(M), and ZnS were measured through IR-ellipsometry, and HfO$_2$ and TiO$_2$ were obtained from ellipsometer database.[39] (c) The optical constants were used to simulate a parameter space with an objective function of Δε and the maximum value is labeled by the symbol, with (d) the simulated spectral emissivity of the cold- and hot- states.

The wavelength-integrated emissivity at temperature $T$ is given by:[11]

$$\varepsilon_T = \frac{\int_{2\mu m}^{18\mu m} \varepsilon(\lambda,T) \, I_B(\lambda,T) \, d\lambda}{\int_{2\mu m}^{18\mu m} I_B(\lambda,T) \, d\lambda} \qquad (1),$$

where $I_B(\lambda, T)$ is the blackbody spectrum. The measured values *n* and *k* of $VO_2$ were used to compute the total emissivity in the insulating (30ºC) and metallic (80ºC) states, along with their difference Δε. According to Kirchoff's law, emissivity in either the hot or cold state is equal to the corresponding absorptivity.

In Figure 2c, Transfer Matrix Method (TMM) parameter sweeps of the $VO_2$ and ZnS thickness were used to generate the bivariate color maps of the difference in emissivity between the cold- and hot- states. To attain the highest Δε, it is advantageous to minimize emissivity in the insulating state, while maximizing it in the metallic state. Denoted with a triangle in the color map, we observe the maximum achievable Δε is ~0.63, and the corresponding simulated spectra is between the states are plotted in Figure 2d. The normalized black body spectra of the corresponding temperatures are also shown to demonstrate the spectra coverage of the simulated emissivity states. Since ZnS has a relatively high refractive index compared to most oxides and even fluorides in the MIR-LWIR, the optimal spacer thickness of 800 nm is lower than for other materials.[14,15]

In this study, we focused on reducing experimental growth time, developing a working prototype, and verifying our results. We thus fabricated samples with varying ZnS thicknesses from 140 nm to 220 nm and a $VO_2$ thickness of ~60 nm. For a ZnS film under 400 nm, our design map (Figure 2e) shows that a thicker $VO_2$ of 50-70 nm can be used with a wide tolerance budget.

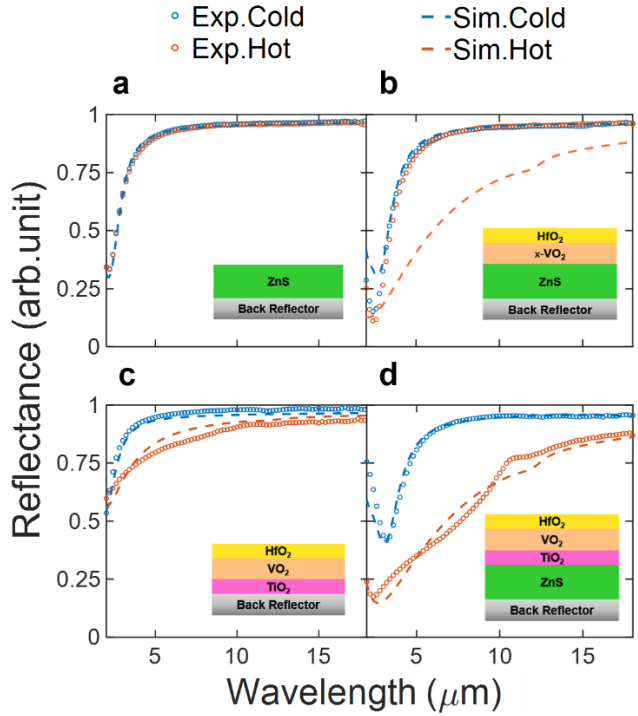

**FIG. 3.** Cold- and hot- state FTIR reflection spectra (unfilled circle) of (a) sample (s1), (b) sample (s3), (c) sample (s4), and (d) sample (s5) are plotted along with their simulated reflection spectra (dash line) in the 2-18 µm wavelength range, using *n* and *k* values from Fig. 2.

We conducted the FTIR spectral measurements using a Bruker Vertex 70 equipped with a Hyperion 3000 microscope system and HgCdTe (MCT) detector. We controlled the temperature using a custom system constructed from a Thorlabs TC300, a Peltier device, and a thermistor. The reflectivity of the cold (30ºC) and hot (80ºC) states for four different samples is shown in Figure 3. Experimental data are shown by symbols, and simulation results are shown by dashed lines. For sample (s1) in Figure 3a, the experimental reflectivity remains virtually unchanged in both states, as ZnS exhibits no phase transition within this temperature range. The remarkable agreement between experimental data and simulations attests to the accuracy of our model. The ZnS reflection valley (around 2 µm) is accurately replicated with TMM, leveraging the measured optical constants shown in Figure 2a. The experimental data for sample (s3), depicted in Figure 3b, exhibits negligible disparity between hot and cold states and diverges from simulated results. This reaffirms our previous assertion that the direct deposition of $VO_2$ onto ZnS fails to induce the monoclinic crystalline structure conducive to phase transition. Figure 3c confirms the presence of phase-change crystalline $VO_2$ grown on $TiO_2$. There is a clear change in the experimental spectrum between cold and hot states, and the spectral characteristics are consistent with simulated results. Subsequently, in Figure 3d, sample (s5)'s reflection intensity exhibits a pronounced decline as it is heated to 80 °C, demonstrating phase change. This experimental data also aligns with our TMM simulation of the entire stack and confirms crystallization of $VO_2$ on the $TiO_2$ buffer layer. Additionally, we can observe the absorption enhancement that the ZnS spacer layer provides by comparing Figure 3c and 3d.

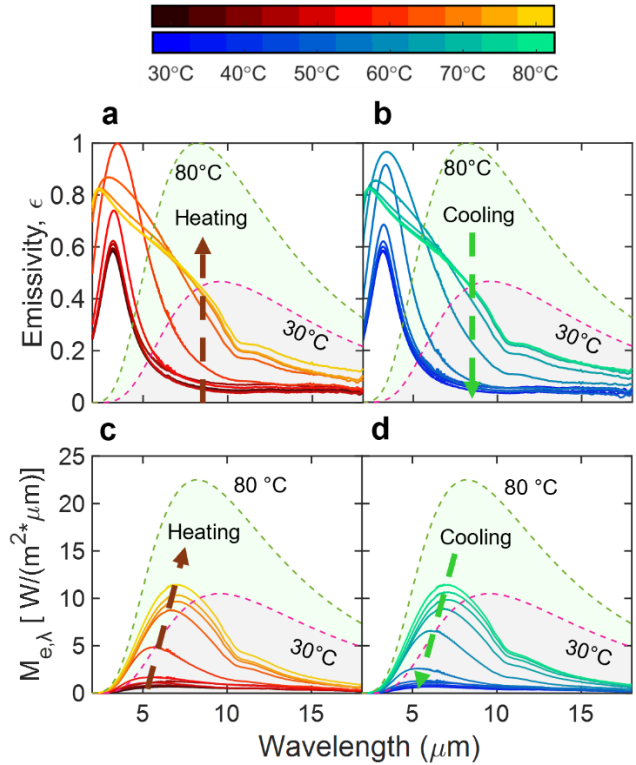

**FIG. 4.** The (a) heating and (b) cooling emissivity spectra of sample (5) is measured at every 5 °C increments is calculated via Eq (1), and shown with the black-body spectrum at 30 °C and 80 °C. The spectral exitance is plotted in (c) and (d).

We next determine the emissivity of our main device, sample (s5), as a function of temperature. From Kirchhoff's law, the spectrally dependent emissivity is equal to the spectrally dependent absorptivity, $\varepsilon(\lambda)=a(\lambda)$. Since our device uses a perfect metal reflector as the bottom layer, we can set $a(\lambda)= 1-R(\lambda)$, where $R$ is reflectivity. These results are shown in Figures 4a and 4b, along with the normalized black body spectra at 30 °C and 80 °C for reference. In Figure 4a, sample (s5) is slowly heated from 30 °C to 80 °C at an interval of 5 °C and is allowed to thermally equilibrate for five minutes before the FTIR reflection spectra is collected. The lines with darker red colors represent lower temperatures, and the lines with lighter orange to yellow colors represent higher temperatures. The spectra show a Fabry-Perot resonance peak, which increases and broadens as $VO_2$ undergoes phase transition. The peak reaches unity absorption at 60 °C in Figure 4a, indicating critical coupling.[40] With further temperature increase, the absorption peak decreases again.

Similarly, the cooling process of sample (s5) is plotted in Figure 4b, with the darker blue colors denoting colder temperatures and the lighter green to turquoise colors denoting hotter temperatures. In both heating and cooling emissivity graphs, we observed no significant emissivity shifts from 30 °C to 50 °C, and then a drastic increase and broadening of the emissivity spectra from 55 °C to 65 °C. The slight difference between the heating and cooling spectra is due to the phase transition hysteresis of $VO_2$ domains.[41]

The spectral exitance, $M_{e,\lambda}$ is defined as the temperature-dependent quantity inside the integral in the numerator of Eq. 1. The spectral exitance for the heating and cooling process is plotted in Figure 4c and 4d. The exitance increases monotonically with temperature upon heating, and it decreases monotonically with cooling. Comparing Figures 4a and 4b with Figures 4c and 4d, we notice that the absorption resonance around 3.5 μm does not significantly affect our spectral exitance, as it lies outside of region in which the blackbody spectrum peaks.

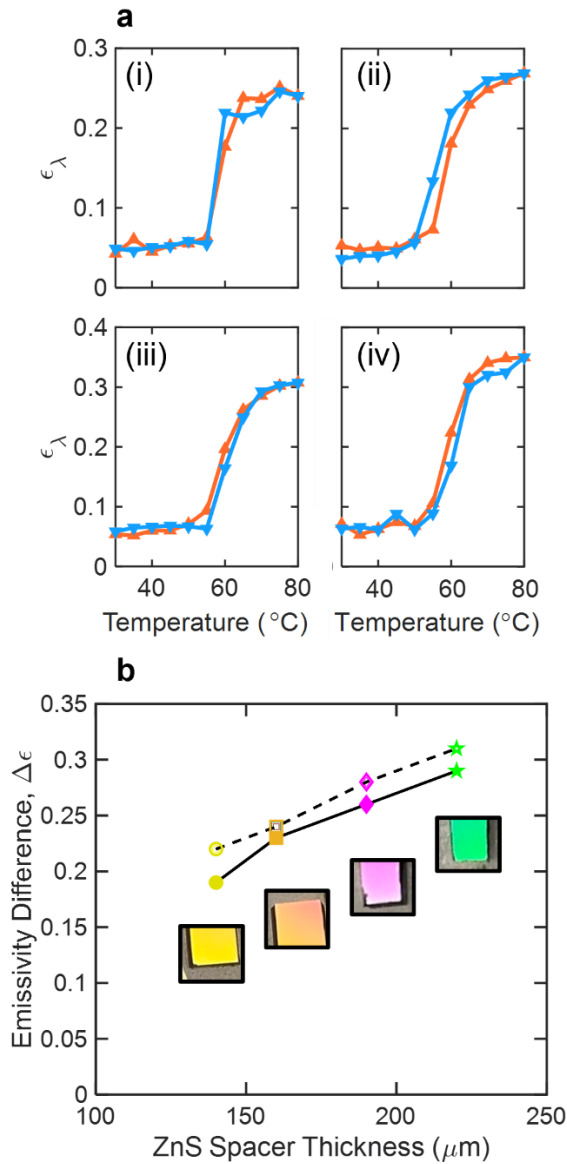

**Figure 5.** The emissivity hystereses (a) of four samples with varying ZnS thickness, (i) 140 nm, (ii) 160 nm, (iii) 190 nm, and (iv) 220 nm are plotted along with their total emissivity difference (b). The experimental (solid line) and simulated (dotted line) results are shown along with insets of the corresponding sample optical images. Each sample data point has the sample color as the corresponding optical image.

We fabricated three additional samples with the same layer composition as sample (s5) and varying spacer layer thickness. The additional devices have ~20 nm $HfO_2$, ~20nm of $TiO_2$, and 60 nm of $VO_2$. The ZnS spacer layer thicknesses of 140 nm, 160 nm, and 190 nm were validated through FTIR and ellipsometry. Following the same procedure above, we collected the FTIR reflectivity data at 5 °C increments from 30 °C to 80 °C and calculated the total spectral emissivity from 2-18 µm. The resulting temperature dependent emissivity is plotted in Figure 5a, in which the red data set represents the heating process, and the blue data set represents the cooling process. Across the devices (i)-(iv), the cold-state emissivity $\varepsilon_c$ remains relatively low (around 0.05). We observe an increasing $\varepsilon_h$ with increasing spacer layer thickness. The emissivity hysteresis in Figure 5a is less than 10ºC, indicating good growth quality. A narrow hysteresis range is pivotal for providing temperature stabilization in response to environmental changes.[10]

The four devices with their measured and simulated $\Delta\varepsilon$ are shown in Figure 5b. Additionally in the figure, inset camera images shows the four different colors produced by ZnS thin film interference. We identified the four samples by their color as Y for yellow, P for peach, M for magenta, and G for green. We observe an increasing trend in $\Delta\varepsilon$ with thicker ZnS spacer layer, in agreement with simulations. These results conclude our successful fabrication and characterization of $VO_2$-ZnS based adaptive thermal emitter.

In summary, we studied the fabrication of adaptive thermal emitters in the $VO_2$-ZnS materials framework using pulsed laser deposition. Direct growth of $VO_2$ on ZnS at 400 °C did not yield any phase change properties in emissivity, indicating the absence of monoclinic $VO_2$. To address this, we introduced an intermediate $TiO_2$ layer to act as a template and buffer between $VO_2$ and ZnS. Devices grown with this intermediary layer exhibited clear $VO_2$ phase change behavior in the temperature range of interest (30 °C to 80 °C). Using the measured $n$ and $k$ values for $VO_2$ and ZnS, we modeled the design space as a function of $VO_2$ and ZnS thickness and found optimal thicknesses with maximum achievable $\Delta\varepsilon$ of 0.63. FTIR measurements conducted on our prototype device demonstrated remarkable agreement with simulations across both cold and hot states. We further fabricated and measured four samples with varying ZnS spacer thickness and demonstrated increasing $\Delta\varepsilon$ in agreement with theoretical predictions. The resulting advancement in passive, thermal regulation schemes could reduce both energy consumption and weight in satellite thermal control systems, as well as other thermal regulation applications.

## ACKNOWLEDGMENTS

This work was funded in part by Northrop Grumman. R. Y. was funded in part by NSF GRFP. The growth chambers used in the study were upgraded with support from AFOSR-DURIP with grant no. FA9550-22-1-0117. The thin film growth of ZnS was supported in part by the ARO under Award No. W911NF-19-1-0137, and an ARO MURI program with award no. W911NF-21-1-0327. The authors acknowledge research group members from Povinelli's lab and Ravichandran's lab for their contribution for making this possible. The author acknowledges the research capabilities provided by the USC's Core Center of Excellence in Nano Imaging and the Raman spectroscopy system from the Cronin Research Lab. The author also thanks Michael Barako, Rachel Rosenzweig, and Max Lien from Northrop Grumman for helpful discussions.

### Data Availability Statement

The data that support the findings of this study are available from the corresponding author upon reasonable request.